\documentclass[aps,twocolumn,groupedaddress,showpacs]{revtex4}
\usepackage{amsmath,amssymb,graphicx,color}
\usepackage{bm}

\begin{document}
\draft
\title{Nonequilibrium work by quantum projective measurements}

\author{Juyeon Yi}

\affiliation{Department of Physics, Pusan National University,
Busan 609-735, Korea}

\author{Yong Woon Kim}

\affiliation{Graduate School of Nanoscience and Technology, \\
Korea Advanced Institute of Science and Technology, Deajeon 305-701, Korea}


\begin{abstract}

We study the thermodynamics of quantum projective measurements by using the set up for the Jarzynski equality.
We prove the fluctuations of energy change induced by measurements satisfy the Jarzynski equality,
revealing that the quantum projective measurements perform a work on a measured system.
When the system is brought into a thermal contact with a reservoir after measurements,
the work done by projective measurements is totally dissipated in the form of heat from the system into the reservoir.
Explicitly showing that the measurements always increase the total entropy,
we prove the link between the second law of thermodynamics and the quantum measurement,
and also provide a clue to the Landau-Lifshitz conjecture on the thermodynamic consequence of measurements.
\end{abstract}


\maketitle

Standard theory of quantum mechanics postulates that upon the action of measurement, the state of a measured system collapses probabilistically onto one of the eigenvectors of the associated observable~\cite{neumann}.
Although its microscopic origin or justification remains to be answered~\cite{namiki, joos2, zurek, schlosshauer}, it is true that this postulate has provided the most precise theory on quantum phenomena.
Still less understood is the thermodynamic consequence of quantum measurement in terms of thermodynamic variables such as work, heat and entropy change. A related issue can be traced back to Landau and Lifshitz who insightfully commented without proof that quantum measurement is a crucial cause for the entropy increase of the second law of thermodynamics~\cite{landau}.
In recent years, using the set up of Maxwell demon, the thermodynamics of quantum measurements has been studied
only in terms of the informational entropy without clear relation to the true thermodynamic entropy~\cite{landauer, bennett,leff,sagawa}.
Also, energetic notions such as heating or cooling were suggested by considering the influence of measurement on the system-reservoir correlation only in specific models~\cite{schulman,erez}.

In this paper, we aim to investigate the thermodynamic nature of quantum measurements without invoking informational entropy and specific models of the system-reservoir coupling.
Here, as a representative quantum measurement, we consider a projective von-Neumann measurement.
We then adopt the set up for the Jarzynski equality~\cite{jarzynski, tasaki, monai}, and derive the work, the heat and the thermodynamic entropy change produced solely by projective quantum measurement. The relation between these thermodynamic quantities is found to guarantee the second law of thermodynamics which determines the direction of the heat flow from the measured system into the reservoir, i.e., we show that the projective measurements perform work on the system, leading to increase of the total entropy.
We for the first time confirm the Landau-Lifshitz conjecture, using the true thermodynamic entropy of thermal equilibrium states,
in the case of the projective measurements.
We also discuss characteristic features of a state of equal probability emergent in the limit of infinite number of measurements~\cite{uniform1, uniform2, uniform3} and quantum Zeno effect~\cite{misra,joo,knight,nakazato,koshino, facchi,zenoexp}, which will be also exemplified by a simple spin model.

It is well worth introducing the Jarzynski equality which is obtained in the set up composed of equilibrium states connected through non-equilibrium processes. As depicted in Fig.~1, initially the system of interest should be prepared in an equilibrium state with a thermal reservoir at temperature $T$. By changing the system Hamiltonian $H(t)$ along a prescribed path~(see the line of $H(t)$ of Fig.~1(a) for $t\in (0,{\cal T}))$, work is performed on the system through the nonequilibrium process. At the end of this work protocol, again brought into the thermal contact with the reservoir at the initial temperature $T$, the system dissipates heat $Q$ into the reservoir and reaches the final equilibrium described by the Hamiltonian $H({\cal T})$ .
Here the work outcome $W$ is in general a stochastic variable and can best be characterized by the probability distribution function $P(W)$. Jarzynski found that the fluctuations of the work are governed by a sum rule~\cite{jarzynski}
\begin{equation}\label{JE}
\int dW P(W) e^{-\beta W}=e^{-\beta \Delta F},
\end{equation}
where $\beta = 1/(k_{B}T)$ and $\Delta F =F_{f}-F_{i}$ with $F_{i}$ and $F_{f}$ denoting the free energy of the initial and the final equilibrium state, respectively.
This remarkable identity relating the average of nonequilibrium quantity to the state function for equilibrium was originally derived for classical system by defining the work as $W=\int_{0}^{{\cal T}}dt\partial H/\partial t$. Later on it is extended to quantum systems\cite{tasaki,monai}, where the work definition is slightly modified as $W=E({\cal T})-E(0)$ with $E(t)$ denoting the energy of the system at a time $t$.

It might have been thought that Jarzynski equality for quantum systems is an extension of its classical version. Yet the problem of quantum measurements provides the drastic difference between the classical and the quantum theory.
If the observable associated with the measurement does not commute with the system Hamiltonian, quantum measurements can yield energy change of the measured system even with the Hamiltonian kept constant in time as sketched in Fig.~1(b), and drive the system out of its initial equilibrium state through a non-unitary evolution. This poses a question, whether this energy change can be entitled to a thermodynamic work satisfying Eq.(1). More importantly, it can further be questioned if any directionality of the work and the heat transfer for the equilibration process would exist in relation to the thermodynamic second law. We will answer these questions using the set up for the Jarzynski equality.

\begin{figure}[b]
\resizebox{8cm}{!}{\includegraphics{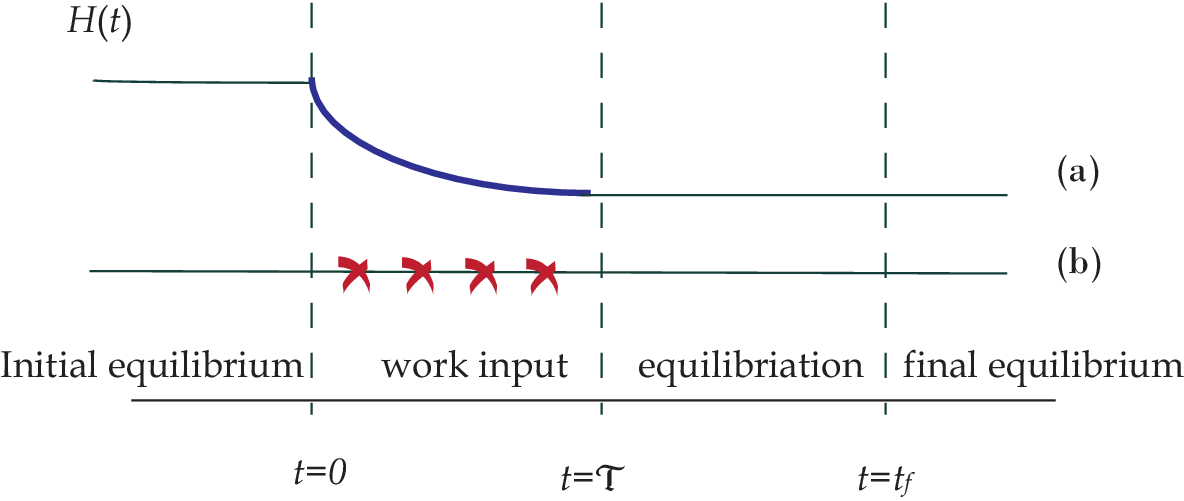}}
 \caption{Schematic diagram of associated stages in Eq.~(1):(a) System is in equilibrium with a thermal reservoir, where the system Hamiltonian is given by $H(0)$ and the corresponding free energy is $F_{i}=-k_{B}T \ln Z$ with $Z=\mbox{Tr}e^{-\beta H(0)}$. For the time span $t\in (0,{\cal T})$, the system is disconnected from the thermal reservoir, and the work is done on the system by changing Hamiltonian from $H(0)$ to $H({\cal T})$ along a prescribed path. If the system with $H({\cal T})$ is then brought in contact with the thermal reservoir, a heat is dissipated into the reservoir, and after sufficiently long time $(t_{f} \gg {\cal T})$ the system and the thermal reservoir achieve the equilibrium state which $F_{f}$ in $\Delta F=F_{f}-F_{i}$ refers to. (b) Only difference from set up in (a) is that the work input is made not by changing the Hamiltonian but by doing projective quantum measurement for $t\in (0,{\cal T})$. Here the system Hamiltonian remains constant but projective measurements interrupting the unitary time evolution of system at the moments marked by crosses bring about the energy change of the sytem.}
\end{figure}

As in the Jarzynski consideration, we suppose that the system of interest is initially in equilibrium
with a thermal reservoir of temperature $T$.
The initial energy of the system, $E(0)$, can be one of the eigenvalues of $H(0)$,  with probability $e^{-\beta E(0)}/Z$, where $Z=\mbox{Tr}e^{-\beta H(0)}$~(Since in our consideration the system Hamiltonian is constant in time, we will write $H(0)=H$ hereafter). Disconnecting then the system from the thermal reservoir, we let the system state evolve in the presence of projective measurements.
We assume that projective measurements are performed every time $\tau$, where the first measurement is done at $t=\tau$ and the last ($M$-th) measurement at ${\cal T}=M\tau$ completes the measurement protocol. At the moment immediately after the projective measurement, the state of the measured system reduces onto one of the eigenstates of an associated observable, say, ${\cal A}=\sum_{\alpha} a_{\alpha}\Pi_{\alpha}$, with the projection operator,
$
\Pi_{\alpha}=|\alpha\rangle \langle \alpha|
$
 and $a_{\alpha}$'s are the eigenvalues of ${\cal A}$: ${\cal A}|\alpha\rangle = a_{\alpha}|\alpha\rangle$. These projectors are assumed to be orthogonal and idempotent,
$
\Pi_{\alpha}\Pi_{\alpha '}=\delta_{\alpha,\alpha'}\Pi_{\alpha},
$
sum to unity
$
\sum_{\alpha}\Pi_{\alpha}=1
$
and can be degenerate
$
\mbox{Tr}\Pi_{\alpha}=g_{\alpha}.
$
The time evolution of the state is then determined by a combined action of the unitary time evolution and projections to give the state vector at the end of the measurement protocol $t={\cal T}$,
\begin{equation}
|\psi;{\cal T}\rangle = \overline{U}_{\{\alpha \}}({\cal T}) |E(0)\rangle,
\end{equation}
where $\{\alpha\}$ represents a set of outcomes from $M$ times measurements, $\{\alpha_{1},\alpha_{2},\cdots,\alpha_{M}\}=\{\alpha\}$. Here the time evolution operator in the presence of $M$ state reductions is given as
\begin{equation}
\overline{U}_{\{\alpha\}}({\cal T})=\Pi_{\alpha_{M}}U(\tau)\cdots \Pi_{\alpha_{2}}U(\tau)\Pi_{\alpha_{1}}U(\tau),
\end{equation}
where the time independent Hamiltonian generates the unitary time evolution operator for a time interval $\tau$ between consecutive projections, $U(\tau)=\exp(-iH\tau/\hbar)$.
The system described by the state $|\psi;{\cal T}\rangle $ can have energy $E({\cal T})$ with a transition probability,
\begin{equation}\label{tran}
\Gamma^{\{\alpha\}}_{{\cal T},0}=|\langle E({\cal T})|{\overline U}_{\{\alpha\}}({\cal T})|E(0)\rangle |^{2}.
\end{equation}
The energy change $W\equiv E({\cal T})-E(0)$ is stochastic for the initial energies distributed according to the equilibrium probability $p(E)=e^{-\beta E}/Z$, and for the transition probability (\ref{tran}) depending on the random outcomes of measurements $\{\alpha\}$.
We note here that so far, $W$ has no thermodynamic meaning other than the energy change by the measurements.
The probability distribution function is then given by;
\begin{equation}\label{pdf1}
P(W)=\sum_{E({\cal T}),E(0)}\sum_{\{\alpha\}}\delta (W-E({\cal T})+E(0))\Gamma^{\{\alpha\}}_{{\cal T},0}p(E(0)),
\end{equation}
where $\sum_{\{\alpha\}}$ denotes the summation over all the possible sequences of the measurement outcomes realized in each running of the protocol.
For this probability distribution function, we can evaluate the left hand side of Eq.~(1) to find,
\begin{equation}\label{eq1}
\int dW e^{-\beta W} P(W) =1,
\end{equation}
where the completeness $\sum_{E(0)}|E(0)\rangle \langle E(0)|=1$ and $\sum_{\{\alpha\}}{\overline U}_{\{\alpha\}}({\cal T}){\overline U}^{\dagger}_{\{\alpha\}}({\cal T}) = 1$ are used.
This explicitly proves that the quantum measurements lead to the energy change which we can identify with a non-equilibrium work,
satisfying the Jarzynski equality.

Note here that in the recent studies~\cite{campisi,rana}, the authors considered a combined situation of protocol (a) and (b) in Fig.~1, where the system dynamics is governed by time varying Hamiltonian as well as by the projective measurement, and proved the invariance of the Jarzynski equality even under the action of projective measurements. The identity, Eq.~(\ref{eq1}) seems obvious in this regard and also tells that $\Delta F =0$, i.e., the quantum measurement alone does not lead to any new equilibrium state of the measured system.
Since the initial and final equilibrium states of the system are thermodynamically identical to each other,
neither internal energy nor the entropy of the system changes after measurements. On the other hand, according to the first law of thermodynamics, the internal energy change between the initial and the final equilibrium is given by $\Delta U=\langle W\rangle -\langle Q\rangle$, where $W$ is the work done on the system for the time elapse $t\in(0,{\cal T})$ and $Q$ is the heat transferred from the system into the reservoir during the equilibration process for $t\in ({\cal T},t_{f})$~(see Fig.~1). Applying the Jensen's inequality $e^{-\beta \langle W\rangle} \leq \langle e^{-\beta W}\rangle = 1$ and using $\Delta U=0$ in this case, we obtain
\begin{equation}\label{ineq}
\langle W\rangle = \langle Q\rangle=T\Delta S_{r}\geq 0 ,
\end{equation}
where the second equality relating to the entropy change of the reservoir $\Delta S_{r}$ holds if
the heat is absorbed by an ideal reservoir in a reversible way.
This relation tells that the average of the work generated by the projective measurement is totally dissipated into the reservoir for the equilibration process in the form of heat. Due to the heat absorption, the entropy of the reservoir increases, indicating the entropy increase of the total system for no entropy change in the system.
This is a proof based on the bona-fide thermodynamic entropy that the Landau-Lifshitz conjecture is indeed correct at least in the case of the projective von-Neumann measurements.
The positivity of the heat indicates the direction of the heat flow from the system into reservoir. This offers an interpretation that the state of the system at the end of repeated projective measurements is {\it hotter} than the reservoir or equivalently initial temperature of the system, although in a strict sense the temperature concept cannot be applied to the nonequilibrium state. There exist some features where we can appreciate the temperature notion of this {\it hot} nonequilibrium state induced by projective measurements such as temperature dependence of $\langle Q\rangle$ and the asymptotic behaviors in the limit of many measurements, as shall be discussed in the following.

The work average can be obtained from the probability distribution Eq.~(\ref{pdf1})
\begin{equation}\label{workave}
\langle W\rangle =\mbox{Tr}H\rho({\cal T})-\mbox{Tr}H\rho(0),
\end{equation}
where the density matrix at the initial equilibrium state is denoted by $\rho(0)=e^{-\beta H}/Z$ and
the density matrix at the time $t={\cal T}$ is given by
\begin{equation}\label{den}
\rho({\cal T}) =\sum_{\{\alpha\}}{\overline U}_{\{\alpha\}}({\cal T})\rho(0){\overline U}^{\dagger}_{\{\alpha\}}({\cal T})
\equiv {\cal M}\rho({\cal T}-\tau).
\end{equation}
Here we define a linear map
\begin{equation}
{\cal M}\rho = \sum_{\alpha}\Pi_{\alpha}U(\tau)\rho U^{\dagger}(\tau)\Pi_{\alpha}.
\end{equation}
Note that if the initial equilibrium state of a system is prepared at infinite temperature~($\beta =0$), the initial density matrix is given by $\rho(0)=(1/D)\openone$ with $\openone$ being the identity matrix and $D$ being the Hilbert space dimension of the system. Since the eigen-projection operators $\Pi_{\alpha}$'s and the unitary time evolution operator $U(\tau)$ commute with $\rho(0)$, we obtain $\rho({\cal T})=\rho(0)$. As a consequence, the work average indeed vanishes $\langle W\rangle =\langle Q\rangle = 0$ at infinite temperature as it should.

Let us remark on the limiting behaviors of the work average depending on the time scales associated with the measurement process. In the limit of many measurements ${\cal T} \gg \tau$, the density matrix $\rho({\cal T})$ should be given to satisfy
$
\rho({\cal T})\approx {\cal M} \rho({\cal T}).
$
By inspection, we can find that $\rho({\cal T})=(1/D)\openone$ is the solution. Similar discussion can be found in Refs.~\cite{uniform1,uniform2}.
The solution is unique if the time duration between measurements $\tau$ is neither zero nor accidently coincident with a special value depending on the details of the system dynamics, for example, the recurrence time as proven in Ref.~\cite{uniform3}.
This state of equal probability for the density matrix taking the same form as infinite temperature yields the maximum work average that can be written in terms of equilibrium internal energy:
\begin{equation}\label{maxwork}
\langle W\rangle = U(0)-U(\beta),
\end{equation}
where $U(\beta)$ is the internal energy of the system at the inverse temperature $\beta$.
On the other hand, when the system is subject to continuous measurements~($\tau \rightarrow 0$),
 the work average even from infinitely many measurements is equivalent to that from a single measurement.
  This can easily be seen by noting that the time evolution operator becomes, $U(\tau)\approx \openone-(iH/\hbar)\tau+(-iH/\hbar)^{2}\tau^{2}$, and further
resulting in the density matrix
\begin{equation}
\rho({\cal T})=\sum_{\alpha}\Pi_{\alpha}\rho(0)\Pi_{\alpha} + {\cal O}(\tau^{2}).
\end{equation}
Neglecting higher orders in $\tau$, this density matrix is identical to that for a single measurement.
This is the manifestation of quantum Zeno effect that continuous measurements totally freeze the quantum state evolution.

We now exemplify the aforementioned behaviors by considering a spin-$1/2$ particle in the presence of a crossed magnetic field $\mathbf{B}=B_{0}{\widehat z}+B_{1}{\widehat x}$. The governing Hamiltonian, $H=-(e/mc)\mathbf{S}\cdot \mathbf{B}$ can be written as
\begin{equation}\label{ham}
H=\epsilon|-\rangle \langle -| -\epsilon|+\rangle \langle +| -\gamma|-\rangle \langle +|
-\gamma |+\rangle \langle -|,
\end{equation}
where $\epsilon = (e\hbar/2mc) B_{0}$ and $\gamma = (e\hbar/2mc) B_{1}$, and $|\pm \rangle$ denote the eigenstates of Pauli spin along $z$ component $\sigma_{z}|\pm\rangle = \pm |\pm \rangle$.
Assume that this system is initially in equilibrium with a reservoir of inverse temperature $\beta$.
For this example we consider the eigen-projectors
\begin{equation}
\Pi_{1}=|-\rangle \langle -| ~~\mbox{and}~~ \Pi_{2}=|+\rangle \langle +|
\end{equation}
for the observable, $S_{z}$, associated with the $z$-component of the spin.
These orthogonal ($\Pi_{1}\Pi_{2}=0$) and idempotent ($\Pi_{\alpha}^{2}=\Pi_{\alpha}$) operators are
complete, $\sum_{\alpha = 1}^{2} \Pi_{\alpha}=1$, and non-degenerate $\mbox{Tr}\Pi_{\alpha}=1$, giving the Hilbert space dimension of the system: $\sum_{\alpha}\mbox{Tr}\Pi_{\alpha}=2$.
A straightforward algebra leads to the Fourier transform of the probability distribution function:
\begin{eqnarray}\label{chatwo}
G(u)&=&\int dW e^{iuW}P(W) \\ \nonumber
&=&\frac{1+\Lambda}{2}+\left[\frac{\cosh(\beta E + 2iuE)}{\cosh(\beta E)}\right]\frac{1-\Lambda}{2},
\end{eqnarray}
where $E=\sqrt{\epsilon^{2}+\gamma^{2}}$.
Putting $u=i\beta$ into Eq.~(\ref{chatwo}), we find that Eq.~(\ref{eq1}) is satisfied. This two level system illustrates that the energy fluctuation caused by projective measurement is governed by the fluctuation theorem, Eq.~(\ref{eq1}),  irrespectively of the system details and the choice of observable.
In Eq.~(\ref{chatwo}), the factor $\Lambda$ depends on $M$ and $\tau$ determining how long and how frequent the projective measurements are performed:
\begin{equation}\label{lambda}
\Lambda = \frac{\epsilon^{2}}{E^{2}}\left[1-2\frac{\gamma^{2}}{E^{2}}\sin^{2}(E\tau/\hbar)\right]^{M-1},
\end{equation}
which comes into play in determining the work average. Taking derivative of Eq.~(\ref{chatwo}) with respect to $u$, $(-i)[\partial G(u)/\partial u ]_{u=0}=\langle W\rangle$, we obtain
\begin{equation}\label{twowork}
\langle W\rangle
=(1-\Lambda) E\tanh (\beta E)
\end{equation}
Unless $\tau$ is an integer multiple of $\pi\hbar/E$, the absolute value of the factor in the square bracket in Eq.~(\ref{lambda}) is less than unity. Therefore, vanishing $\Lambda$ in the limit $M\rightarrow \infty $ yields
$\langle W\rangle = E\tanh (\beta E)$. Noting that the internal energy of the system at the inverse temperature $\beta$ is given by $U(\beta)=-E\tanh(\beta E)$, one can find that this work average corresponds to the internal energy difference of the system at infinite temperature and at the initial temperature, as shown in Eq.~(\ref{maxwork}). On the other hand, for the continuous measurement with $\tau =0$, we have $\Lambda = \epsilon^{2}/E^{2}$, independently of $M$. The corresponding work average
is given by $\langle W\rangle \approx (\gamma^{2}/E)\tanh(\beta E)$ which is identical to the work average for a single measurement~($M=1$), exhibiting the quantum Zeno effect. Temperature dependence of $\langle W\rangle$ can explicitly be appreciated in this example. For a reservoir at very high temperature~($\beta E \ll 1$), Eq.~(\ref{twowork}) gives vanishingly small work average $\langle W\rangle \approx (1-\Lambda)E^{2}\beta$, for which the reservoir absorbs negligible amount of heat and its entropy change is also very small as $\Delta S_{r} \sim (\beta E)^{2}$. On the other hand, the work average in the low temperature regime~($\beta E \gg 1$) is saturated into $\langle W\rangle \approx (1-\Lambda)E$, which leads to the entropy change of the reservoir as $\Delta S_{r} \sim \beta E$.

Any two level systems can realize this example and yet the experimental verification of our theory crucially depends on designing two different measurements: One is the energy measurement for the work $W$ and the other should be associated with an observable noncommuting with the system Hamiltonian. Recent experiments in cold atom physics have shown the possibility to perform spin measurements through optical control~\cite{np}. In particular, the energy measurement by using spectroscopy for trapped cold ions demonstrated a way to confirm the quantum Jarzynski equality~\cite{lutz}. With all implied possibilities of experimental realizations of the measurement protocols, we would like to again emphasize that experimental confirmation of Eq.~(6) must directly reveals the thermodynamic consequence of quantum measurements as the nonequilibrium work and the thermodynamic entropy production.



We are grateful to Peter Talkner for helpful discussions. 
Y.W.K. acknowledges support from Basic Science Research Program through the National Research Foundation of Korea (NRF) funded by the Ministry
of Education, Science and Technology(Grant No. 2010-0025196).
J.Y. acknowledges support from the National Research Foundation of Korea (NRF) grant funded by the Korea government (MEST) (Grant No.  2011-0021296).

\end{document}